# Fourier spectra for nonuniform phase-shifting algorithms based on principal component analysis

Manuel Servin,* Moises Padilla, Guillermo Garnica, and Gonzalo Paez

*Centro de Investigaciones en Optica A. C., Loma del Bosque 115, Lomas del Campestre, 37150 Leon Guanajuato, Mexico.*
*mservin@cio.mx*

**Abstract**: We develop an error-free, nonuniform phase-stepping algorithm (nPSA) based on principal component analysis (PCA). PCA-based algorithms typically give phase-demodulation errors when applied to nonuniform phase-shifted interferograms. We present a straightforward way to correct those PCA phase-demodulation errors. We give mathematical formulas to fully analyze PCA-based nPSA (PCA-nPSA). These formulas give a) the PCA-nPSA frequency transfer function (FTF), b) its corrected Lissajous figure, c) the corrected PCA-nPSA formula, d) its harmonic robustness ($R_H$), and e) its signal-to-noise-ratio (SNR). We show that the PCA-nPSA can be seen as a linear quadrature filter, and as consequence, one can find its FTF. Using the FTF, we show why plain PCA often fails to demodulate nonuniform phase-shifted fringes. Previous works on PCA-nPSA (without FTF), give specific numerical/experimental fringe data to "visually demonstrate" that their new nPSA works better than competitors. This often leads to biased/favorable fringe pattern selections which "visually demonstrate" the superior performance of their new nPSA. This biasing is herein totally avoided because we provide figures-of-merit formulas based on linear systems and stochastic process theories. However, and for illustrative purposes only, we provide specific fringe data phase-demodulation, including comprehensive analysis and comparisons.

## 1. Introduction

Phase-shifting interferometry is a widely used optical metrology technique to demodulate the phase from a set of phase-shifted interferograms [1–7]. The demodulated phase contains the searched measuring information from *N* phase-shifted fringe patterns. Traditionally phase shifting algorithms (PSAs) require precise phase-shifting steps, however, it is not easy to be absolutely sure that an interferometer have a zero-error phase-shifter. Therefore, methods to estimate nonuniform phase-shifting steps and the desired modulating phase from nonlinear phase-shifted data have been investigated [8–23]. The Lissajous ellipse fitting technique is one of the earliest phase demodulation methods for dealing with nonuniform phase-stepped images [24–26]. If the demodulated analytic signal form a Lissajous circle, one obtains an error-free measuring phase [24-26]. For erroneous phase demodulation, the Lissajous figure of the analytic signal is not a circle, it is an ellipse. The ellipse fitting method convert the Lissajous ellipse into a Lissajous circle, and the phase demodulation error is eliminated [27-32]. The Lissajous technique is a powerful technique to correct phase demodulation errors; it is currently based on least squares fitting of a rotated and origin-shifted ellipse; this has however its own difficulties which are sometimes not trivially solved [27–32].

   Another nonuniform phase-steps algorithm (nPSA) is the principal component analysis (PCA) of phase-shifted fringe data [33–38]; we call this the PCA-nPSA technique. This is a subspace technique because it finds two orthogonal signals from *N* correlated nonuniform phase-shifted fringe images. Unmodified (or plain) PCA-nPSA demodulate the phase from fringe images without the explicit knowledge about their nonlinear phase shifts. The PCA-nPSA technique has low computational cost, it is linear, it is non-iterative, and it can deal with spatially varying background illumination and fringe contrast. Therefore, it seems at first glance, that PCA-nPSA would deal with all possible situations of nonuniform/linear phase-shifted phase demodulation [33-38]..

In spite of all those good properties, the PCA-nPSA has however some disadvantages which often gives inacceptable phase demodulation errors [37-40]. The PCA-nPSA users may not be aware of the phase-demodulation errors, and may therefore reach erroneous conclusions in phase metrology engineering [37-40]. A well studied PCA-nPSA limitation occurs when less-than-one spatial fringe is present within the fringe pattern [33–40]. But the problem of "few-spatial-fringes" is in our view, a pseudo-problem. That is because one can easily introduce as many spatial fringes as desire simply by introducing a large spatial carrier (a large tilt), and the few-spatial-fringes "problem" is gone [1]. A more recent attempt, and good review, to improve the PCA-nPSA using the Lissajous figure is given in [40].

In this work we show that the PCA-nPSA can be regarded as a linear quadrature filter applied to nonuniform phase-shifted fringe data. And as any other linear filter, it is possible to find its Fourier spectrum through its frequency transfer function (FTF) [1]. Finding the FTF of the PCA-nPSA one can see the reason why plain PCA often fails to demodulate, error-free, a set of nonuniform phase-shifted data. With the FTF at hand, one can find the signal-to-noise ratio (SNR) and fringe harmonics robustness of the PCA-nPSA from first principles of stochastic process and linear systems theories [1].

## 2. Nonuniform phase-shifting fringe images

We first describe the continuous-time fringe model as,

$$I(x,y,t) = a(x,y) + b(x,y)\cos[\varphi(x,y) + \omega_0 t]; \quad t \in \mathbb{R}. \tag{1}$$

Where $a(x,y)$, $b(x,y)$, and $\varphi(x,y)$ are the background, the amplitude, and the phase of the fringes respectively. The parameter $\omega_0$ is the angular frequency of the fringes. Without loss of generality we assume $\omega_0=1.0$ (radians/second). The temporal Fourier transform $F[\cdot]$ of the fringe is,

$$F[I(t)] = a\delta(\omega) + \frac{b}{2}e^{-i\varphi}\delta(\omega+1) + \frac{b}{2}e^{i\varphi}\delta(\omega-1). \tag{2}$$

For all $(x,y) \in L \times L$, and $i = \sqrt{-1}$; see Fig. 1.

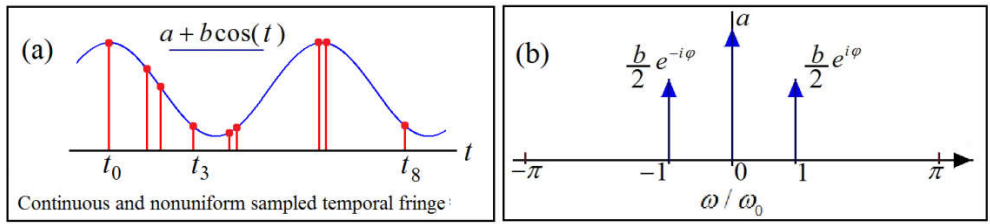

Fig. 1. Panel (a) shows 9 nonuniform temporal samples of a continuous-time fringe. Panel (b) shows the Fourier spectrum of the continuous-time fringe.

The temporal continuous fringe in Eq. (1) is sampled at nonuniform times $(t_n)$ as,

$$I_n = \int_{-\infty}^{\infty} [a + b\cos(\varphi + \omega_0 t)]\delta(t-t_n)dt; \quad n \in \{0,1,...,N-1\}. \tag{3}$$

It is usual to label the fringe samples by their nonlinear phase-steps $\theta_n = \omega_0 t_n$ as,

$$I_n = a + b\cos(\varphi + \theta_n); \quad \forall (x,y) \in L \times L. \tag{4}$$

Note that $\omega_0 = 1.0$ and $t_n$ are not meaningful; the relevant data are the nonuniform phase-steps $\theta_n = \omega_0 t_n$. Figure 1(a) shows 9 nonlinearly-spaced phase-shifted samples (in red), and Fig. 1(b) the Fourier spectrum of the continuous temporal fringe (in blue).

### 3. PCA-nPSA phase demodulation formula

Principal component analysis (PCA) was invented by Karl Pearson in 1901 [42] and it is a statistical procedure that uses a linear transformation that convert hundreds of correlated observations, into a subset of linearly uncorrelated signals called the principal components of the data. In phase-shifting demodulation, the PCA is used to find 2 orthogonal signals (an analytic signal) from few temporal fringe samples. The reason of using a handful (instead of hundreds) of nonuniform phase-shifted data makes that the PCA-nPSA often give erroneous phase estimation, making the PCA-nPSA (if not properly corrected) inadequate for precision optical metrology. That is why several works have been published to correct the PCA-nPSA to cope with this residual phase demodulation error [37-41].

We now construct the desired PCA-nPSA formula. We start by modeling $N$ nonuniform phase-shifted samples as,

$$I_n(x,y) = a(x,y) + b(x,y)\cos[\varphi(x,y) + \theta_n]; \quad n = \{0,1,...,N-1\}. \tag{5}$$

The images have $(x,y) \in L \times L$ pixels. Firstly we estimate the background of the fringes as,

$$\hat{a}(x,y) = \frac{1}{N}\sum_{n=0}^{N-1} I_n(x,y). \tag{6}$$

The following step in PCA is to compute the covariance matrix,

$$[\mathbf{C}]_{m,n} = \frac{1}{L^2}\sum_{x=0}^{L-1}\sum_{y=0}^{L-1}[I_m(x,y) - \hat{a}(x,y)][I_n(x,y) - \hat{a}(x,y)]. \tag{7}$$

We now find the $N$ eigenvalues $\lambda_n$ and eigenvectors $\mathbf{v}_n$ of matrix $\mathbf{C}$ as,

$$\mathbf{C}\mathbf{v}_n = \lambda_n \mathbf{v}_n; \quad n \in \{0,1,...,N-1\}. \tag{8}$$

Assuming that the largest eigenvalues are $(\lambda_0, \lambda_1)$ ($\mathbf{C}\mathbf{v}_0 = \lambda_0 \mathbf{v}_0$, $\mathbf{C}\mathbf{v}_1 = \lambda_1 \mathbf{v}_1$), then the PCA-nPSA formula is given by,

$$A(x,y) = \sum_{n=0}^{N-1}\left([\mathbf{v}_0]_n + i[\mathbf{v}_1]_n\right) I_n(x,y). \tag{9}$$

Where $A(x,y)$ is the demodulated analytic signal, and its phase is given by $\arg[A(x,y)]$. Note that the PCA as herein presented do not need to vectorize back-and-forth the fringe images [33-36]. Equation (9) is the searched non-corrected (plain) PCA-nPSA formula. Equation (9) constitutes an interesting result because however implicit in the original PCA technique [33], it has not been explicitly given as an standard phase-shifting algorithm formula (Eq. (9)).

The first work on PCA as nPSA was presented as a linear algorithm which could demodulate any set of phase-shifted fringes, almost error-free [33]. Afterwards this was found not to be exact and several attempts have been made to improve plain PCA [35,38-40]. However, the PCA can be combined with AIA, obtaining a PCA-AIA algorithm which eliminate the phase-errors left by plain PCA [35]. The PCA providing the first phase estimation making the AIA to converge faster [35]. Note that the AIA estimate both, the modulating phase $\varphi(x,y)$ and the nonuniform phase-steps $\{\theta_0, \theta_1, ..., \theta_{N-1}\}$ [35].

### 4. Correcting the Lissajous ellipse of the PCA-nPSA analytic signal

The Lissajous figure is obtained by the following parametric plot,

$$\mathbf{r}[\varphi(x,y)] = \mathrm{Re}[A(x,y)]\,\mathbf{i} + \mathrm{Im}[A(x,y)]\,\mathbf{j}. \tag{10}$$

Where **i** and **j** are the real and imaginary unit vectors. A good advantage of PCA-nPSA is that it always give zero-rotated Lissajous ellipses $\mathbf{r}[\varphi]$. Zero-rotated ellipses are easily transformed into circles (with zero phase demodulation error) by first calculating the ratio,

$$\rho = \frac{\sum_{x=0}^{L-1}\sum_{y=0}^{L-1}|\mathrm{Im}[A(x,y)]|}{\sum_{x=0}^{L-1}\sum_{y=0}^{L-1}|\mathrm{Re}[A(x,y)]|}\;; \qquad (0 < \rho \leq 1.0). \tag{11}$$

Where $|x|$ denotes the absolute value of *x*. Equation (11) is very robust to noise because a single parameter is estimated from an entire image. With $\rho$, we may transform the Lissajous ellipse into a circle by modifying plain PCA-nPSA (Eq. (9)) as,

$$A_2(x,y) = \sum_{n=0}^{N-1}\left(\rho[\mathbf{v}_0]_n + i[\mathbf{v}_1]_n\right)I_n(x,y). \tag{12}$$

Now we have a Lissajous circle, *i.e.* an error-free phase $\arg[A_2(x,y)]$. Equation (12) constitutes the corrected PCA-nPSA formula. The Lissajous ellipses are then transformed into circles without explicitly knowing the nonuniform phase-steps $\{\theta_0, ..., \theta_{N-1}\}$ [27-31].

### 5. Frequency transfer function (FTF) for the PCA-nPSA

The phase-steps $\{\theta_0, ..., \theta_{N-1}\}$ can be known by combining the PCA and AIA (PCA-AIA) algorithms [35]. Having $\{\theta_0, ..., \theta_{N-1}\}$ we can obtain the FTF from the PCA-nPSA formula because it may be seen as a convolution product (see [1]),

$$A(x,y) = \sum_{n=0}^{N-1} c_n^* I_n(x,y) = [h(t) * I(x,y,t)]_{t=N-1},$$

$$h(t) = \sum_{n=0}^{N-1} c_n \delta\left(t - \frac{\theta_n}{1.0}\right);\quad c_n = \rho[\mathbf{v}_0]_n + i[\mathbf{v}_1]_n\,; \tag{13}$$

$$I(x,y,t) = \sum_{n=0}^{N-1} I(x,y,\theta_n)\delta\left(t - \frac{\theta_n}{1.0}\right).$$

Where $h(t)$ is the impulse response and $I(x,y,t)$ is the fringe data. Coefficients $c_n^*$ are the complex conjugate of $c_n$, and linear convolution is represented by ($*$) [1]. The Fourier transform of $h(t)$ is the FTF of the PCA-nPSA filter, or $H(\omega) = F[h(t)]$ as,

$$H(\omega) = F\left[\sum_{n=0}^{N-1} c_n \delta\left(t - \frac{\theta_n}{1.0}\right)\right] = \sum_{n=0}^{N-1} c_n e^{-i\theta_n\omega}. \tag{14}$$

To obtain a quadrature signal, $H(\omega)$ must comply at least, with the following conditions,

$$H(-1) = 0; \qquad H(0) = 0; \qquad H(+1) \neq 0. \tag{15}$$

The zero $H(-1) = 0$ rejects the conjugate signal $(b/2)e^{-i\varphi}\delta(\omega+1)$, and $H(0) = 0$ rejects the background $ae^{-i\varphi}\delta(\omega)$. Then one obtains an error-free analytic signal as,

$$A(x,y) = \frac{b(x,y)}{2}H(1)e^{i\varphi(x,y)}. \tag{16}$$

If the response at $H(-1)$ is not zero, then one obtains an erroneous analytic signal given by,

$$A_{error} = \frac{b}{2}\left[e^{-i\varphi}H(-1) + e^{i\varphi}H(1)\right] = \frac{b}{2}H(1)e^{i\varphi}\left[1 + \frac{H(-1)}{H(1)}e^{-2i\varphi}\right]; \quad \forall (x,y) \in L \text{x} L. \tag{17}$$

A non-zero $H(-1)$ generate a detuning-like phase-demodulation error $[H(-1)/H(1)]e^{-2i\varphi}$. This is the typical phase error given by plain PCA-nPSA.

## 6. Signal-to-noise ratio gain ($G_{SNR}$) and fringe harmonic robustness

Once the FTF (Eq. (14)) is obtained, one can find the SNR and harmonics robustness of the PCA-nPSA from basic stochastic process and linear systems theories [1]. Without the FTF people usually rely on particular synthetic/experimental fringe images (sometimes favorably biased) which may lead to over-optimistic conclusions [9-40].

### 6.1 Signal-to-noise ratio gain $G_{SNR}$ of the PCA-nPSA formula

The SNR of the analytic signal (Eq. (13)) for nonuniform sampled fringes corrupted by additive white Gaussian noise (AWGN) with power density $N(\omega) = \eta/2$, $\omega \in (-\pi, \pi)$ is [1].

$$\text{SNR} = \frac{Quadrature\ Signal\ Power}{Filtered\ \text{AWGN}\ Power} = \frac{(b^2/4)|H(1.0)|^2}{(\eta/2)\sum_{n=0}^{N-1}|c_n|^2}. \tag{18}$$

From this formula we define the SNR-gain ($G_{SNR}$) as [1],

$$G_{SNR} = \frac{|H(1.0)|^2}{\sum_{n=0}^{N-1}|c_n|^2}; \quad (0 < G_{SNR} \leq N). \tag{19}$$

The number $G_{SNR}$ is the SNR of the analytic signal with respect to the SNR of the fringe data. For example $G_{SNR} = N$ means that the analytic signal has $N$-times higher SNR than the fringe data. The number $G_{SNR}$ reduces substantially for highly nonuniform phase-step fringes [41].

### 6.2 Fringe harmonic robustness $R_H$ for N-steps PCA-nPSA

Phase-shifted harmonic-distorted fringes may be modeled by,

$$I(\varphi, \theta_n) = a + b\cos[\varphi + \theta_n] + \sum_{k=2}^{\infty} b_k \cos[k\varphi + k\theta_n]; \quad \forall (x,y) \in L \text{x} L. \tag{20}$$

Then the demodulated analytic signal $A(x,y)$ for harmonic distorted fringes is given by,

$$A(x,y) = \frac{b}{2}e^{i\varphi}H(1) + \sum_{k=2}^{\infty}\left(\frac{b_k}{2}\right)\left[e^{ik\varphi}H(k) + e^{-ik\varphi}H(-k)\right]. \tag{21}$$

Where $H(k)$ and $H(-k)$ is the FTF response to the $k$-th harmonics $k=\{2,3,...\}$. Assuming that the harmonics amplitude decreases as $b_k=(1/k)$ we define the harmonic robustness as,

$$R_H = \frac{Quadrature\ Signal\ Power}{Total\ Fringe\ Harmonic\ Power} = \frac{|H(1.0)|^2}{\sum_{k=2}^{\infty}\left(\frac{1}{k^2}\right)\left[|H(k)|^2 + |H(-k)|^2\right]}. \quad (22)$$

A large $R_H$ number means high fringe harmonics robustness. In contrast, a low $R_H$ means low fringe harmonics robustness.

### 7. Computer simulations

For illustrative purposes only, we now offer two simulation for plain and corrected PCA-nPSA applied to 3 and 9 nonuniform phase sampled fringe images. These examples are given to show the Fourier spectral response of plain/corrected PCA-nPSA.

#### 7.1 Plain PCA-nPSA applied to 3 nonuniform phase-step fringe images

The PCA-nPSA has been applied to many temporal fringe samples for better results [33-36]. That is because PCA was conceived to extract uncorrelated signals from hundreds correlated statistical data. The PCA was not invented to extract an analytic signal from, let say, 5 fringe samples. However the herein corrected PCA-nPSA, can demodulate the phase from just 3 temporal samples.

Let us start with the 3 nonuniform phase-shifted fringe images shown in Fig. 2 and Fig. 3.

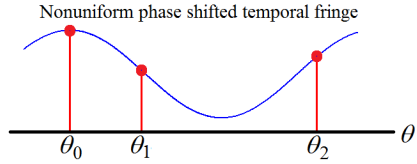

Fig. 2 Three nonlinear phase-shifted samples taken at $\theta_n=\{0, 1.49, 5.13\}$ radians.

The nonlinear phase-shifting are $\theta_n = \{0, 1.49, 5.13\}$ radians. The fringes are shown in Fig. 3.

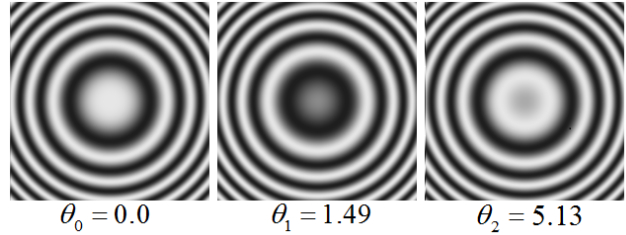

$\theta_0 = 0.0$ $\quad\theta_1 = 1.49\quad$ $\theta_2 = 5.13$

Fig. 3. The 3 nonuniform phase-shifted fringe images.

Then we apply the PCA to these 3 images as,

$$[C_{n,m}] = \begin{bmatrix} 3.89 & -3.73 & -0.16 \\ -3.73 & 17.02 & -13.3 \\ -0.16 & -13.3 & 13.46 \end{bmatrix} \Rightarrow \begin{pmatrix} \lambda_0 \\ \lambda_1 \\ \lambda_2 \end{pmatrix} = \begin{pmatrix} 28.95 \\ 5.42 \\ 0 \end{pmatrix} \quad (23)$$

Taking the largest eigenvalues $\mathbf{C}\mathbf{v}_0 = \lambda_0 \mathbf{v}_0$ and $\mathbf{C}\mathbf{v}_1 = \lambda_1 \mathbf{v}_1$, the PCA-nPSA is,

$$A(x,y) = \sum_{n=0}^{2} \left([\mathbf{v}_0]_n + i[\mathbf{v}_1]_n\right) I_n(x,y). \qquad (24)$$

The demodulated phase $\arg[A(x,y)]$ is given in Fig. 4 along with its phase error.

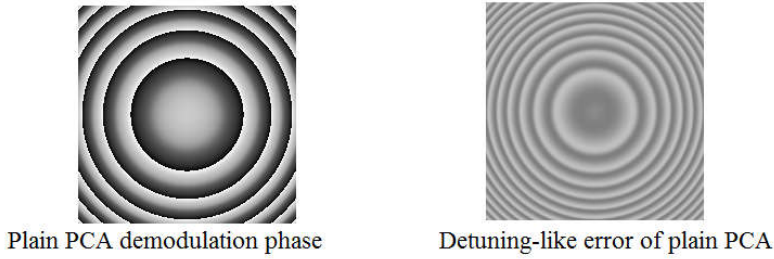

Fib. 4. Here we show plain PCA-nPSA demodulated phase, and its phase-error.

The FTF $H(\omega)$ of this (plain) PCA-nPSA is,

$$H(\omega) = \sum_{n=0}^{2} \left([\mathbf{v}_0]_n + i[\mathbf{v}_1]_n\right) e^{-i\theta_n \omega}. \qquad (25)$$

The FTF of plain PCA-nPSA is plotted in Fig. 5(a). We see that $|H(-1)| > 0$ and the erroneous analytic signal is given by,

$$A = \frac{b}{2} H(1) e^{i\varphi} \left[1 + \frac{H(-1)}{H(1)} e^{-2i\varphi}\right] = \frac{b}{2} e^{i\varphi} \left[1 + 0.31 e^{-2i\varphi}\right]; \quad \forall (x,y). \qquad (26)$$

From Fig. 5, $H(1) = 1.0$, and $H(-1) = 0.31$, giving an error of $0.31 e^{-2i\varphi}$.

### 7.2 Lissajous-figure for 3 nonuniform phase-steps PCA-nPSA

Figure 5(a) shows the FTF, analytic signal, and Lissajous ellipse of $A(x,y)$ (Eq. (24)),

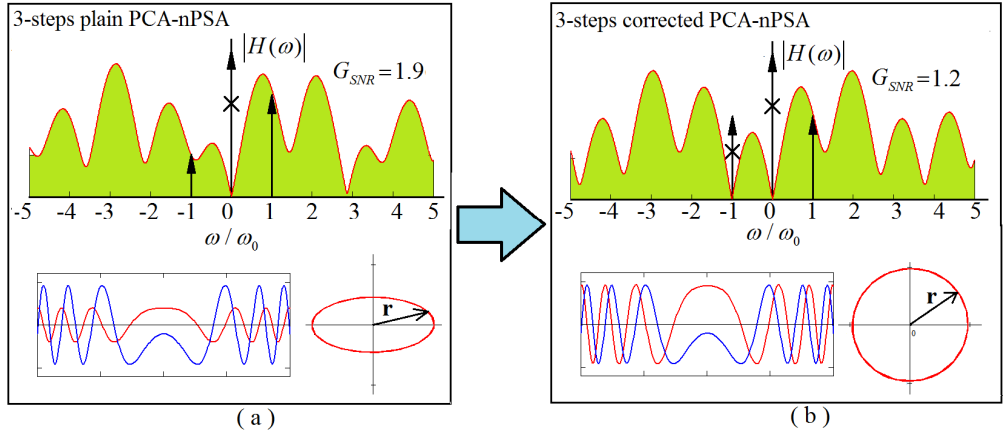

Fig. 5. Correcting 3-steps plain PCA. Panel (a) shows the FTF, analytic signal, and Lissajous ellipse of plain PCA. Panel (b) shows the corrected FTF along its analytic signal and Lissajous circle. We also show the SNR-gain slight degradation due to FTF correction.

The noise-robust, correcting factor $\rho$ is,

$$\rho = \frac{\sum_{x=0}^{L-1}\sum_{y=0}^{L-1}\left|\operatorname{Im}[A(x,y)]\right|}{\sum_{x=0}^{L-1}\sum_{y=0}^{L-1}\left|\operatorname{Re}[A(x,y)]\right|} = 0.432. \tag{27}$$

Figure 5(a) shows plain PCA's FTF, its analytic signal and its Lissajous ellipse. We then use $\rho$ to correct the PCA-nPSA obtaining the Lissajous circle in Fig. 5(b). From Fig. 5 we see that $G_{SNR}$ reduces from 1.96 to 1.2. On the other hand, the harmonic robustness $R_H$ decreases from 1.3 to 0.66. That is, the corrected PCA-nPSA is more sensitive to noise and harmonics than plain PCA-nPSA. However the phase error of plain PCA-nPSA is intolerable (see Fig. 4).

### *7.3 Correction of plain PCA-nPSA applied to 9 nonuniform phase-step fringe data*

We now phase demodulate 9 nonlinear phase-shifted fringe images; Fig. 9 shows 4 images,

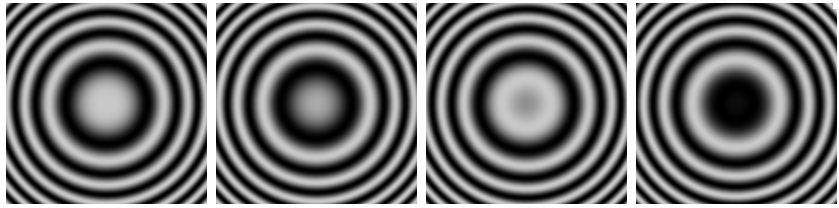

Fig. 6. Here we show 4 out of 9 nonlinear phase-shifted fringe patterns.

Figure 7 shows the 9 phase-steps $\theta_n = \{0, 1.13, 2.49, 1.52, 3.55, 3.78, 6.2, 6.42, 8.74\}$,

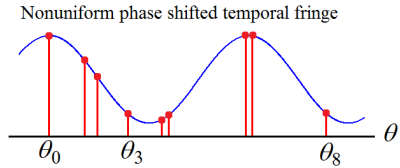

Fig. 7. The red dots plot represent 9 nonlinear/nonuniform phase steps..

Figure 8(a) shows the FTF of plain PCA and Fig. 8(b) the corrected FTF,

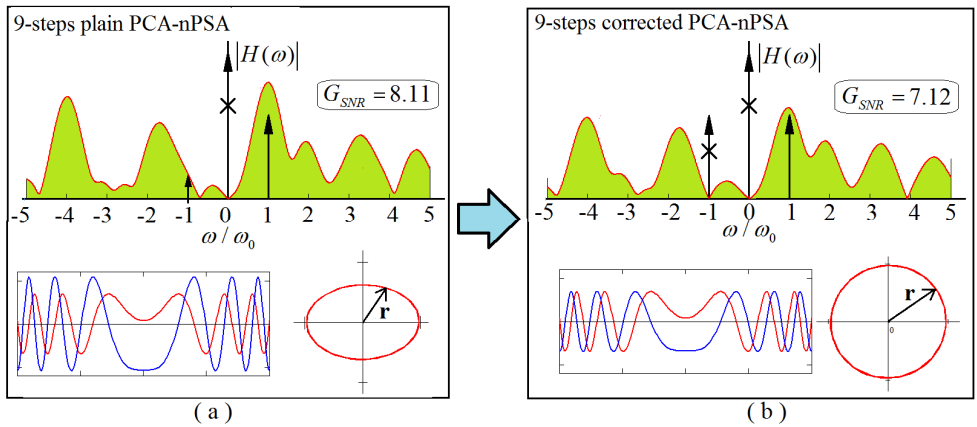

Fig. 8. Correcting 9-steps plain PCA. Panel (a) shows the FTF, analytic signal, and Lissajous ellipse of plain PCA applied to Fig. 7 data. Panel (b) shows the corrected FTF, analytic signal, and Lissajous circle. We also show the SNR-gain slight degradation due to FTF correction.

The SNR-gain $G_{SNR}$=8.11 is higher for plain PCA-nPSA, but the phase-error is intolerable. The harmonic robustness for plain PCA-nPSA is $R_H$=4.316, while for the *corrected* PCA-nPSA is $R_H$=3.381. Plain PCA-nPSA has better harmonics robustness. Figure 8 also shows the Lissajous ellipse for plain PCA and the Lissajous circle for the corrected PCA-nPSA.

## 8. Conclusion

We have presented a very simple way to correct the technique of principal component analysis (PCA) applied to phase-demodulation of nonuniform phase-shifting fringes. We can summarize the contributions of this work as,

- We have presented a PCA-nPSA procedure which do not need to vectorize back and forth the nonuniform phase-sampled fringe images (Eqs. (5)-(9)).

- Applying the PCA-nPSA formula we found that the Lissajous figures of the demodulated analytic signal are always non-rotated ellipses.

- The non-rotated Lissajous ellipses are corrected to Lissajous circles using the corrected PCA-nPSA formulas given in Eq. (11) and Eq. (12).

- With the PCA-nPSA formula, and knowing the phase steps $\{\theta_0, \theta_1, ..., \theta_{N-1}\}$, one can find the PCA-nPSA Fourier spectral response $H(\omega)$ (the FTF).

- The FTF of the PCA-nPSA is then used to estimate the SNR-gain ($G_{SNR}$) Eq. (19) from basic stochastic processes theory for fringes corrupted by AWGN [1].

- Also the FTF ($H(\omega)$) is used to estimate the harmonics robustness $R_H$ (Eq. (22)) of the plain/corrected PCA-nPSA formulas.

- We have finally shown that plain PCA-nPSA has better SNR and harmonics robustness ($R_H$) than corrected PCA-nPSA. But plain PCA-nPSA has, in general, inacceptable phase demodulation errors.

In brief, Fig. 5 and Fig. 8 show that the price for correcting plain PCA's demodulated phase are: 1) a small increase in SNR, and 2) a small decrease in harmonics robustness.

**References**

1. M. Servín, J. A Quiroga, and M. Padilla, *Fringe Pattern Analysis for Optical Metrology*, (WILEY-VCH, 2014).
2. H. Schreiber and J. H. Bruning, "Phase shifting interferometry," in Optical Shop Testing (Wiley, 2006), pp. 547–666.
3. Y.-Y. Cheng and J. C. Wyant, "Multiple-wavelength phase-shifting interferometry," Appl. Opt. **24**, 804–807 (1985).
4. H. Medecki, E. Tejnil, K. A. Goldberg, and J. Bokor, "Phase-shifting point diffraction interferometer," Opt. Lett. **21**, 1526–1528 (1996).
5. I. Yamaguchi and T. Zhang, "Phase-shifting digital holography," Opt. Lett. **22**, 1268–1270 (1997).
6. Y. Awatsuji, M. Sasada, and T. Kubota, "Parallel quasi-phase-shifting digital holography," Appl. Phys. Lett. **85**, 1069–1071 (2004).
7. K. Ishikawa, K. Yatabe, N. Chitanont, Y. Ikeda, Y. Oikawa, T. Onuma, H. Niwa, and M. Yoshii, "High-speed imaging of sound using parallel phase-shifting interferometry," Opt. Express **24**, 12922–12932 (2016).
8. G. Lai and T. Yatagai, "Generalized phase-shifting interferometry," J. Opt. Soc. Am. A **8**, 822–827 (1991).
9. K. Okada, A. Sato, and J. Tsujiuchi, "Simultaneous calculation of phase distribution and scanning phase shift in phase shifting interferometry," Opt. Commun. **84**, 118–124 (1991).
10. I-B. Kong and S.-W. Kim, "General algorithm of phase-shifting interferometry by iterative least-squares fitting," Opt. Eng. **34**, 183–188 (1995).
11. Z. Wang and B. Han, "Advanced iterative algorithm for phase extraction of randomly phase-shifted interferograms," Opt. Lett. **29**, 1671–1673 (2004).
12. A. Patil and P. Rastogi, "Approaches in generalized phase shifting interferometry," Opt. Lasers Eng. **43**, 475–490 (2005).
13. H. Guo, Y. Yu, and M. Chen, "Blind phase shift estimation in phase shifting interferometry," J. Opt. Soc. Am. A **24**, 25–33 (2007).
14. X. F. Xu, L. Z. Cai, Y. R. Wang, X. F. Meng, W. J. Sun, H. Zhang, X. C. Cheng, G. Y. Dong, and X. X. Shen, "Simple direct extraction of unknown phase shift and wavefront reconstruction in generalized phase-shifting interferometry: algorithm and experiments," Opt. Lett. **33**, 776–778 (2008).

15. P. Gao, B. Yao, N. Lindlein, K. Mantel, I. Harder, and E. Geist, "Phase shift extraction for generalized phase-shifting interferometry," Opt. Lett. **34**, 3553–3555 (2009).
16. J. Deng, H. Wang, D. Zhang, L. Zhong, J. Fan, and X. Lu, "Phase shift extraction algorithm based on Euclidean matrix norm," Opt. Lett. **38**, 1506–1508 (2013).
17. H. Guo and Z. Zhang,"Phase shift estimation from variances of fringe pattern differences," Appl. Opt. **52**, 6572–6578 (2013).
18. R. Juarez-Salazar, C. Robledo-Sánchez, C. Meneses-Fabian, F. Guerrero-Sánchez, and L. A. Aguilar, "Generalized phase-shifting interferometry by parameter estimation with the least squares method," Opt. Lasers Eng. **51**, 626–632 (2013).
19. J. Deng, L. Zhong, H. Wang, H. Wang, W. Zhang, F. Zhang, S. Ma, and X. Lu, "1-Norm character of phase shifting interferograms and its application in phase shift extraction," Opt. Commun. **316**,156–160 (2014).
20. Q. Liu, Y. Wang, J. He, and F. Ji, "Phase shift extraction and wavefront retrieval from interferograms with background and contrast fluctuations," J. Opt. **17**, 025704 (2015).
21. J. Deng, D. Wu, K. Wang, and J. Vargas, "Precise phase retrieval under harsh conditions by constructing new connected interferograms," Sci. Rep. **6**, 24416 (2016).
22. Y. Xu, Y. Wang, Y. Ji, H. Han, and W. Jin, "Three-frame generalized phase-shifting interferometry by a Euclidean matrix norm algorithm," Opt. Lasers Eng. **84**, 89–95 (2016).
23. X. Xu, X. Lu, J. Tian, J. Shou, D. Zheng, and L. Zhong, "Random phase-shifting interferometry based on independent component analysis," Opt. Commun. **370**, 75–80 (2016).
24. K. Kinnstaetter, A. W. Lohmann, J. Schwider, and N. Streibl, "Accuracy of phase shifting interferometry," Appl. Opt. **27**, 5082–5089 (1988).
25. C. T. Farrell and M. A. Player, "Phase step measurement and variable step algorithms in phase-shifting interferometry," Meas. Sci. Technol. **3**, 953–958 (1992).
26. C. T. Farrell and M. A. Player, "Phase-step insensitive algorithms for phase-shifting interferometry," Meas. Sci. Technol. **5**, 648–654 (1994).
27. A. Albertazzi Jr., A. V. Fantin, D. P. Willemann, and M. E. Benedet, "Phase maps retrieval from sequences of phase shifted images with unknown phase steps using generalized N-dimensional Lissajous figures—principles and applications," Int. J. Optomechatron. **8**, 340–356 (2014).
28. C. Meneses-Fabian and F. A. Lara-Cortes, "Phase retrieval by Euclidean distance in self-calibrating generalized phase-shifting interferometry of three steps," Opt. Express **23**, 13589–13604 (2015).
29. F. Liu, Y. Wu, and F. Wu, "Correction of phase extraction error in phase-shifting interferometry based on Lissajous figure and ellipse fitting technology," Opt. Express **23**, 10794–10807 (2015).
30. F. Liu, Y. Wu, F. Wu, and W. Song, "Generalized phase shifting interferometry based on Lissajous calibration technology," Opt. Lasers Eng. **83**, 106–115 (2016).
31. F. Liu, J. Wang, Y. Wu, F. Wu, M. Trusiak, K. Patorski, Y. Wan, Q. Chen, and X. Hou, "Simultaneous extraction of phase and phase shift from two interferograms using Lissajous figure and ellipse fitting technology with Hilbert Huang prefiltering," J. Opt. **18**, 105604 (2016).
32. A. V. Fantin, D. P. Willemann, M. E. Benedet, and A. G. Albertazzi, "Robust method to improve the quality of shearographic phase maps obtained in harsh environments," Appl. Opt. **55**, 1318–1323 (2016).
33. J. Vargas, J. A. Quiroga, and T. Belenguer, "Phase-shifting interferometry based on principal component analysis,"Opt. Lett. **36**, 1326–1328 (2011).
34. J. Vargas, J. A. Quiroga, and T. Belenguer, "Analysis of the principal component algorithm in phase-shifting interferometry," Opt. Lett. **36**, 2215–2217 (2011).
35. J. Vargas, C. Sorzano, J. Estrada, and J. Carazo, "Generalization of the principal component analysis algorithm for interferometry," Opt. Commun. **286**, 130–134 (2013)..
36. J. Vargas and C. Sorzano, "Quadrature component analysis for interferometry," Opt. Lasers Eng. **51**, 637–641 (2013).
37. J. Xu, W. Jin, L. Chai, and Q. Xu, "Phase extraction from randomly phase-shifted interferograms by combining principal component analysis and least squares method," Opt. Express **19**, 20483–20492 (2011).
38. J. Deng, K. Wang, D. Wu, X. Lv, C. Li, J. Hao, J. Qin, and W. Chen, "Advanced principal component analysis method for phase reconstruction," Opt. Express **23**, 12222–12231 (2015).
39. K. Yatabe, K. Ishikawa, and Y. Oikawa, "Improving principal component analysis based phase extraction method for phase shifting interferometry by integrating spatial information," Opt. Express **24**, 22881–22891 (2016).
40. K. Yatabe, K. Ishikawa, and Y. Oikawa, " Simple, flexible, and accurate phase retrieval method for generalized phase-shifting interferometry," JOSA A. **34**, 6017–6024 (2016).
41. M. Servin, M. Padilla, G. Garnica, and G. Paez, "Design of nonlinear spaced phase-shifting algorithms using their frequency transfer function," Appl. Opt., **58**, 1134-1138 (2019).
42. K. Pearson. "On Lines and Planes of Closest Fit to Systems of Points in Space". Philosophical Magazine. 2 (11): 559–572 (1901).